\documentclass{emulateapj}
\usepackage{graphicx}
\usepackage{times,psfig}

\def\ltsima{$\; \buildrel < \over \sim \;$}
\def\lsim{\lower.5ex\hbox{\ltsima}}
\def\gtsima{$\; \buildrel > \over \sim \;$}
\def\gsim{\lower.5ex\hbox{\gtsima}}
\def\mdot {\dot M}
\newcommand{\be}{\begin{equation}}
\newcommand{\en}{\end{equation}}

\def\kms  {\rm \ km \, s^{-1}}

\def\cmdue {\rm \ cm^{-2}}

\def\msole {~M_{\odot}}

\begin{document}

\title{Outliers from the mainstream: how a massive star can produce a gamma--ray burst}  

\author{S. Campana\altaffilmark{1}, N. Panagia\altaffilmark{2,3,4}, D. Lazzati\altaffilmark{5},
A. P. Beardmore\altaffilmark{6}, G. Cusumano\altaffilmark{7},
O. Godet\altaffilmark{6}, G. Chincarini\altaffilmark{8,1}, S. Covino\altaffilmark{1},
M. Della Valle\altaffilmark{9,10,11}, C. Guidorzi\altaffilmark{8,1},
D. Malesani\altaffilmark{12}, A. Moretti\altaffilmark{1}, R. Perna\altaffilmark{5},
P. Romano\altaffilmark{8,1}, G. Tagliaferri\altaffilmark{1}}

\altaffiltext{1}{INAF - Osservatorio Astronomico di Brera, Via Bianchi 46, I-23807
Merate (Lc), Italy}
\altaffiltext{2}{STScI, 3700 San Martin Drive, Baltimore, MD 21218, USA}
\altaffiltext{3}{INAF - Osservatorio Astrofisico di Catania, via S. Sofia 78,
I-95123 Catania, Italy}
\altaffiltext{4}{Supernova Ltd., Olde Yard Village 131, Northsound Road, Virgin
Gorda, British Virgin Islands}
\altaffiltext{5}{JILA, Campus Box 440, University of Colorado, Boulder, CO 80309-0440,
USA}
\altaffiltext{6}{Department of Physics and Astronomy, University of Leicester, University
Road, Leicester LE1 7RH, UK}
\altaffiltext{7}{INAF - Istituto di Astrofisica Spaziale e Fisica Cosmica di Palermo, via
U. La Malfa 153, I-90146 Palermo, Italy}
\altaffiltext{8}{Universit\`a degli studi di Milano Bicocca, piazza delle Scienze 3,
I-20126 Milano, Italy}
\altaffiltext{9}{European Southern Observatory, Karl-Schwarzschild-Strasse 2
D-85748 Garching bei M\"unchen, Germany}
\altaffiltext{10}{INAF - Osservatorio Astronomico di Capodimonte, salita Moiariello 16,
I-80131 Napoli, Italy}
\altaffiltext{11}{International Center for Relativistic Astrophysics Network, Piazza
della Repubblica 10, I-65122, Pescara, Italy}
\altaffiltext{12}{Dark Cosmology Centre, Niels Bohr Institute, University of Copenhagen,
Juliane Maries vej 30, DK-2100 K\o{}benhavn \O, Denmark}



\begin{abstract}
It is now recognized that long-duration Gamma--Ray Bursts (GRBs) are
linked to the collapse of massive stars, based on the association between
(low-redshift) GRBs and (type Ic) core-collapse supernovae (SNe).
The census of massive stars and GRBs reveals, however, that not all massive
stars do produce a GRB. Only $\sim 1\%$ of core collapse SNe are able to
produce a highly relativistic collimated outflow, and hence a GRB.
The extra crucial parameter has long been suspected to be metallicity and/or
rotation. We find observational evidence
strongly supporting that both ingredients are necessary in
order to make a GRB out of a core-collapsing star. A detailed study
of the absorption pattern in the X--ray spectrum of GRB060218
reveals evidence of material highly enriched in low atomic number
metals ejected before the SN/GRB explosion. We find that, within
the current scenarios of stellar evolution, only a progenitor star
characterized by a fast stellar rotation and sub-solar initial metallicity
could produce such a metal enrichment in its close surrounding.


\end{abstract}

\section{Introduction}

The association between long GRBs and SNe hints toward Wolf-Rayet (WR) stars
as GRB progenitors (Woosley \& Bloom 2006; Fruchter et al. 2006).
The WR phase in the evolution of a massive star is relatively short,
therefore WR stars are rarely observed. They eject high velocity
($v_w\sim 1000-2000\kms$), mass-loaded winds (mass-loss rates of
$\mdot\sim 10^{-5}-10^{-4}\msole$ yr$^{-1}$) as well as massive shells
during mass-loss episodes (e.g. Pastorello et al. 2006; Weiler et al. 2007).
Signatures for the presence of this material can be expected in the optical
light curve of GRB afterglows showing up as flux enhancements in their light
curves or as (variable) fine-structure transition lines in their optical
spectra. The observations of these features can set
important constraints on the density and distance of the absorbing material
located  either in the star-forming region within which the progenitor formed,
or in the circumstellar environment of the progenitor itself (Perna \& Loeb
1998; Prochaska et al. 2005). 

In the X--ray domain, despite the wealth of features produced by absorption of
metals, relatively little progress has been achieved, mostly due to low
statistics and to the relatively poor spectral resolution of the detectors.
Here, we consider the Swift (Gehrels et al. 2004) observations of
GRB060218. This is the second closest GRB (redshift $z=0.033$), and the first
one showing the shock break out of the SN (Campana et al. 2006).
Modelling of the spectra and light curve of the associated SN 2006aj (Pian et
al. 2006; Mirabal et al. 2006; Sollerman et al. 2006; Cobb et al. 2006)
suggested a progenitor star whose initial mass was $20\pm1 \msole$  
(Mazzali et al. 2006).  

This GRB was of very long duration, which allowed Swift to observe
it with its narrow field instruments (the X--ray Telescope, XRT, Burrows et
al. 2005, and the UV/Optical Telescope, UVOT) during a considerable part of 
its prompt phase, collecting the largest number of X--ray photons ever.
In this letter we exploit the huge number of X--ray photons by analysing the
absorption pattern burnt into it by circumburst material.

\section{X--ray data analysis}

The XRT spectra have been obtained in the Windowed Timing (WT) mode in which a
1D image is obtained by adding the data along the central 200 pixels in
a single row (see Hill et al. 2004).
The XRT data have been processed using the FTOOLS software package (v.~6.3.1)
distributed within HEASOFT. We run the task {\em xrtpipeline}
(v.0.11.4) applying calibration and standard filtering and screening
criteria. In particular, we dynamically correct for possible bias offsets
computing the bias difference between the on-ground estimated bias median from
the last 20 pixels data telemetered with every frame, and the median of the
last 20 pixels in the related bias row.
Events with grade 0 have been selected in order to attain the best
spectral resolution. The XRT analysis has been performed in the 0.3--10 keV
energy band (and also in the 0.35--10 keV and 0.25--10 keV energy bands as a
consistency check). Given the XRT CCD resolution at low
energies ($\Delta E/E\sim 15\%$) we cannot directly see the edges imprint in
the X--ray spectrum. Rather we are sensitive to the different slopes in
between the edges leading to a precise determination of the depths of the
single edges.

We extract the data from an 80 pixel wide region, given the strength of the
source. The background has been extracted at the edge of the WT slit accounting
for only $\sim 0.5\%$ of the source flux. A dedicated arf file has been generated
with the {\tt xrtmkarf} task accounting for bad column holes and correcting
for vignetting and point spread function losses. The latest response matrices
have been used (v.010, see Campana et al. 
2006\footnote{http://heasarc.gsfc.nasa.gov/docs/heasarc/caldb/swift/docs/xrt/SWIFT-XRT-CALDB-09}).
These matrices represent a big improvement with respect to the previous
version. Calibration data on Mkn421 and isolated neutron stars indicate a good
response down to 0.3 keV with no apparent features and data to model ratios
always below $10\%$. A systematic uncertainty of $2.5\%$ has been estimated
for very bright sources, at the most.

We divide the entire GRB light curve into two segments taking the XRT peak
as dividing line in order to minimize the effects of spectral
variations (Campana et al. 2006). This choice results into two intervals:
the first 715 s (count rate 72 c s$^{-1}$, for a total of 51,000 photons) and
the second 1907 s long (73 c s$^{-1}$, for a total of 139,000 photons).
The peak rate is 130 c s$^{-1}$, well below the WT pile-up limit.
Data has been rebinned to have at least 100 counts per energy bin.

For the same intervals we extract survey mode BAT data (Barthelmy et al. 2005)
in order to fit the XRT and BAT spectra simultaneously and better constrain
the XRT high energy part of the emission model. The XRT energy range is
0.3--10 keV and the BAT range is 16--79 keV and 16--37 keV for the first and
second (softer) spectrum, respectively. 

The large number of counts allowed us for the first time to quantify the
abundances of a number of elements, especially those absorbing at low
energies. The most prominent absorption edges are those caused by carbon,
nitrogen, oxygen, neon, magnesium, silicon, sulfur, iron and nickel (Morrison
\& McCammon 1983). 
We considered a simple absorption model including absorption
by gas in our Galaxy and intrinsic absorption in the host galaxy (at the known
redshift, both modelled with {\tt tbabs}, Wilms et al. 2000, and the standard
XSPEC abundance pattern, Anders \& Grevesse 1989) folding a black body plus a
cut-off power law emission model (the soft component is mandatory to obtain a
good fit at variance with most GRBs, Campana et al. 2006).  
The Galactic absorption has been left free
to vary within the interval $(9.8-15)\times 10^{21}$ cm$^{-2}$. This has been
estimated by accumulating 360 ks over the field of GRB060218 and fitting the
spectra of three relatively bright sources ($\sim 5\times 10^{-14}$ erg
s$^{-1}$ cm$^{-2}$) with more than 200 counts with the same absorbing column
density. The fit of the three sources with a power law is satisfactory
and defines the $90\%$ confidence interval as above. This range is also
consistent with the column density estimate from HI maps (Dickey \&
Lockman 1990; Kalberla et al. 2005) and dust NIR maps converted into X--ray
absorption (Schlegel et al. 1998). The intrinsic column density is free to
vary at the host galaxy redshift of $z=0.03342$ (Wiersema et al. 2007). 
We also considered the inclusion of a gain fit (a small rigid
adjustment of the energy scale due to residuals in the bias subtraction) and a
systematic uncertainty at the $2.5\%$ level. 

Adopting a solar metallicity for the host galaxy absorption provides a rather
poor fit with large residuals in the 0.4--0.6 keV energy range, as expected (see
Fig. 1). We then left free the metallicity of the host galaxy. We obtained a
better fit with a lower metallicity (using the {\tt tbvarbas} model)
$Z=0.20^{+0.02}_{-0.06}\,Z_\odot$ (hereon uncertainties are given at the
$90\%$ confidence level) and a large column  density of $N_{\rm
HI}=(2.0^{+0.7}_{-0.4})\times10^{22}\cmdue$. The metallicity is higher than
the mean value determined for the host galaxy
($Z=0.07^{+0.06}_{-0.02}\,Z_\odot$, see Wiersema et al. 2007). In addition,
the (solar composition) HI column density is so high that, even with that low
metallicity, it would imply an extinction of a factor of $\gsim 50$ higher
than what has been inferred from the strengths of the sodium absorption lines
in the host galaxy ($E(B-V)=0.042\pm0.003$, Wiersema et al. 2007) or from the
Balmer decrement ($E(B-V)\lsim 0.03$, Guenther et al. 2006). The improvement
from a solar metallicity absorber to a free metallicity absorber is at the
$4.2\,\sigma$ level (by means of an F-test).  A similar discrepancy between
optical and X--ray absorption is frequently observed in GRB afterglows
(Galama \& Wijers 2001; Stratta et al. 2005; Watson et al. 2007) even adopting
the dust to gas ratio of the Small Magellanic Cloud  (SMC, Pei et al. 1992).


Despite the large number of counts, we do not have enough sensitivity
to leave free the abundances of all individual elements in the fitting
procedure. We focus therefore on the CNO abundances
(with absorption edges at rest-frame energies of 0.28, 0.41 and 0.54
keV, respectively) as additional free parameters, while we adopt the
abundances derived from optical studies for the remaining
elements. The CNO abundances are (much) larger than the solar value. 
Fitting for a gain offset further improves the fit with values in the --30 to --10 eV
range. With this model we obtain a better fit to the data with an
F-test probability of 5.1--5.5$\,\sigma$ (depending on the adopted
systematic uncertainties in the response matrix). 
The host galaxy column density is in this case $N_{\rm HI}=1.1^{+0.9}_{-0.4}
\times10^{21}\cmdue$, which is consistent with the optical absorption measured
in the host galaxy for an SMC dust to gas ratio. This result may indicate that
the discrepancy between optical and X--ray absorption is simply due to a large
amount of circumburst material enriched in CNO elements. If these are mostly  
in the form of gas then they cause absorption in the X--rays but not in the
optical.

The error on the CNO abundances has been evaluated with the {\it steppar}
command within XSPEC to circumvent non-monotonicity in $\chi^2$ space.
We also tried a different absorption model using {\tt varabs} within XSPEC,
obtaining very similar results.

An additional issue could be the rapidly evolving spectrum, leading to
spurious spectral features. To test whether this could pose a problem, we
divide the two initial spectra into six smaller intervals each. We generate
the appropriate response and then fit with the same model as above (and
absorption parameters fixed) the resulting spectra. Then, we simulate spectra
with a factor of 10 more photons and join them. The two resulting
fake spectra recover quite well the initial absorption pattern (now left free
to vary) within the fitting uncertainties.

GRB060218 occurred in February 2006. Data has been processed with the
appropriate version 6 gain files ({\tt swxwtgains0\_20010101v006.fits},
calibrated in 2005). As a test, we also used version 7 of the gain file ({\tt
swxwtgains0\_20010101v007.fits}, calibrated in late 2006--2007). This new gain
file together with a new arf file (yet unreleased) provides a superior
description of the spectral models but it is fully appropriate for
observations taking place nowadays. We reprocess all the data and fit
them with the same best fit model. We check that also in this case we obtain
the same results well within the uncertainties and, more importantly, the same
ratios by number. This positive test further strengthens our observational
result.


\bigskip

\section{Discussion and conclusions}

Since we do not have a knowledge of the total mass involved, the most
insightful measurement we obtain from X--ray best fit data is the ratio by
number of the abundances of C, N and O. We derive ${\rm
C/N}=0.9^{+3.2}_{-0.7}$ and ${\rm O/N}=0.3^{+0.6}_{-0.2}$ (or logarithmic
abundances relative to the solar value $[{\rm C/N}]=-0.5^{+0.7}_{-1.5}$ and
$[{\rm O/N}]=-1.4^{+0.5}_{-0.2}$). This extremely low ratio O/N is very
difficult to account for in terms of standard interstellar medium and of
stellar evolution models of isolated stars. In particular, solar metallicity
models are unable to reproduce these abundance ratios (Hirschi et al. 2005;
Portinari et al. 1998). Possibly, binary evolutionary models can account for
this constraints more easily, having more degrees of freedom, but recent
models seem to indicate that conditions similar to single star progenitors are
needed (Detmers et al. 2008).

A key ingredient in the evolution of single massive stars is missing:
rotation-induced mixing in the stellar interior. 
If the ejected mass reflects the C/N and O/N ratios that would be expected at
the end of the main sequence phase as observed in several nebulae around bright
stars, we would need a large rotation-induced mixing fraction with only $\sim
10\%$ of the initial mass unprocessed (based on calculations in Lamers et
al. 2001). This mixing can occur only in the case of a very fast stellar
rotation (close to break-up). A viable alternative could be also provided by a
close binary system, either in terms of tidal locking or evolution through a
common envelope phase during which an enriched shell might be ejected. 

The collapsar scenario requires massive helium stars with rapidly
rotating cores to produce a GRB (Woosley 1993).
However, stellar models with magnetic torques fail to retain such high core
angular momentum. In the last few years there has been mounting theoretical
support the idea that only massive stars that are initially very rapidly
rotating and have sufficiently low metallicities can satisfy the conditions
for GRB formation (Yoon \& Langer 2005; Woosley \& Heger 2006). In fact, below
a suitable metallicity threshold, a rotationally-induced mixing process
produces a quasi chemically homogeneous stellar evolution avoiding the
spin-down of the stellar core. As a test-bed we consider massive star
evolution models at sub-solar initial metallicities (Yoon et al. 2006). 
We also limit the mass range of the progenitor to $15-25\msole$, according to
the detailed modelling of the supernova ejecta (Mazzali et al. 2006).
We find that a number of models are able to satisfy our constraints (see Fig. 2). All
these models are characterized by a fast semi-convective mixing of the core.
Within the initial mass range of the progenitor we are able to
constrain the initial fraction of the Keplerian velocity ($v_{\rm K}$) of the
equatorial rotational velocity to $0.45\lsim v_{\rm ini}/v_{\rm K}\lsim 0.8$
and the initial metallicity to $Z<0.1\,Z_\odot$.
With these parameters the progenitor star fits nicely within the
allowed region for the GRB production (Yoon et al. 2006).
It thus appears that the observations of GRB060218 provide the first
observational evidence that only a progenitor star characterized by a fast
stellar rotation and sub-solar initial metallicity can lead to such an
explosive event.

\begin{acknowledgements}
We acknowledge S.-C. Yoon for help with Fig. 2 and D.N. Burrows for comments. 
SC thanks Dr. G. Balza, E. Biguzzi, M. Tavola for making this letter possible. 
We acknowledge partial support from ASI (ASI/I/R/039/04) and PPARC. The Dark 
Cosmology Centre is supported by the Danish National Research Foundation.
\end{acknowledgements}

\begin{table}
\caption{Spectral fitting results with different energy ranges and fit options.}
\begin{tabular}{cccccc}
\hline
Model & Energy range & Gain & System. & $\chi^2_{\rm red}$ &F-test$^*$  \\
      & (keV)        &      &($2.5\%$)& (dof)$^+$  [nhp$^+$ $\%$]  &($\sigma$) \\
\hline
1$^\#$& 0.25--10     & N    & N       & 1.11 (830) [0.01]  & --      \\
2$^\#$& 0.30--10     & N    & N       & 1.10 (827) [0.02]  & --      \\
3$^\#$& 0.35--10     & N    & N       & 1.10 (825) [0.02]  & --      \\
\hline
4     & 0.3--10      & Y    & N       & 1.07 (825) [0.06]  & 5.5      \\
5     & 0.3--10      & Y    & Y       & 0.96 (825) [81.]   & 5.1      \\
\hline
\end{tabular}

$^+$ dof: degrees of freedom; nhp: null hypothesis probability.

$^*$ Improvement in the fit by leaving free to vary the abundances of CNO,
evaluated by means of an F-test.

$^\#$ The enlargement of the energy band produces a worsening of the $\chi^2$,
indicating that the low energy part of the considered interval plays an
important role in the evaluation of the goodness of fit.

\end{table}

\begin{table*}[!hb]
\begin{center}
\caption{Spectral fit absorption parameters.}
\label{obslog}
\begin{tabular}{cccccc}
\hline
Element & Edge energy & Edge energy     & Solar abundance& Abundance&  Column density\\
        & (keV)       & rest-frame (keV)& table (H$=1$)   & ($Z_\odot$)&  (cm$^{-2}$) \\
\hline
C       &0.268 & 0.277                  & $3.6\times 10^{-4}$ & $15^{+48}_{-10}$  & $6^{+19}_{-4}\times10^{18}$ \\
N       &0.380 & 0.392                  & $1.1\times 10^{-4}$ & $52.5^{+28}_{-12}$& $7^{+4}_{-2}\times10^{18}$\\
O       &0.508 & 0.525                  & $8.3\times 10^{-4}$ & $2.2^{+0.3}_{-0.2}$& $2^{+0.2}_{-0.3}\times10^{18}$\\
\hline
\end{tabular}
\end{center}
\end{table*}

\newpage

\begin{figure*}[!htbp]
\vskip 10truecm
\begin{center}
\psfig{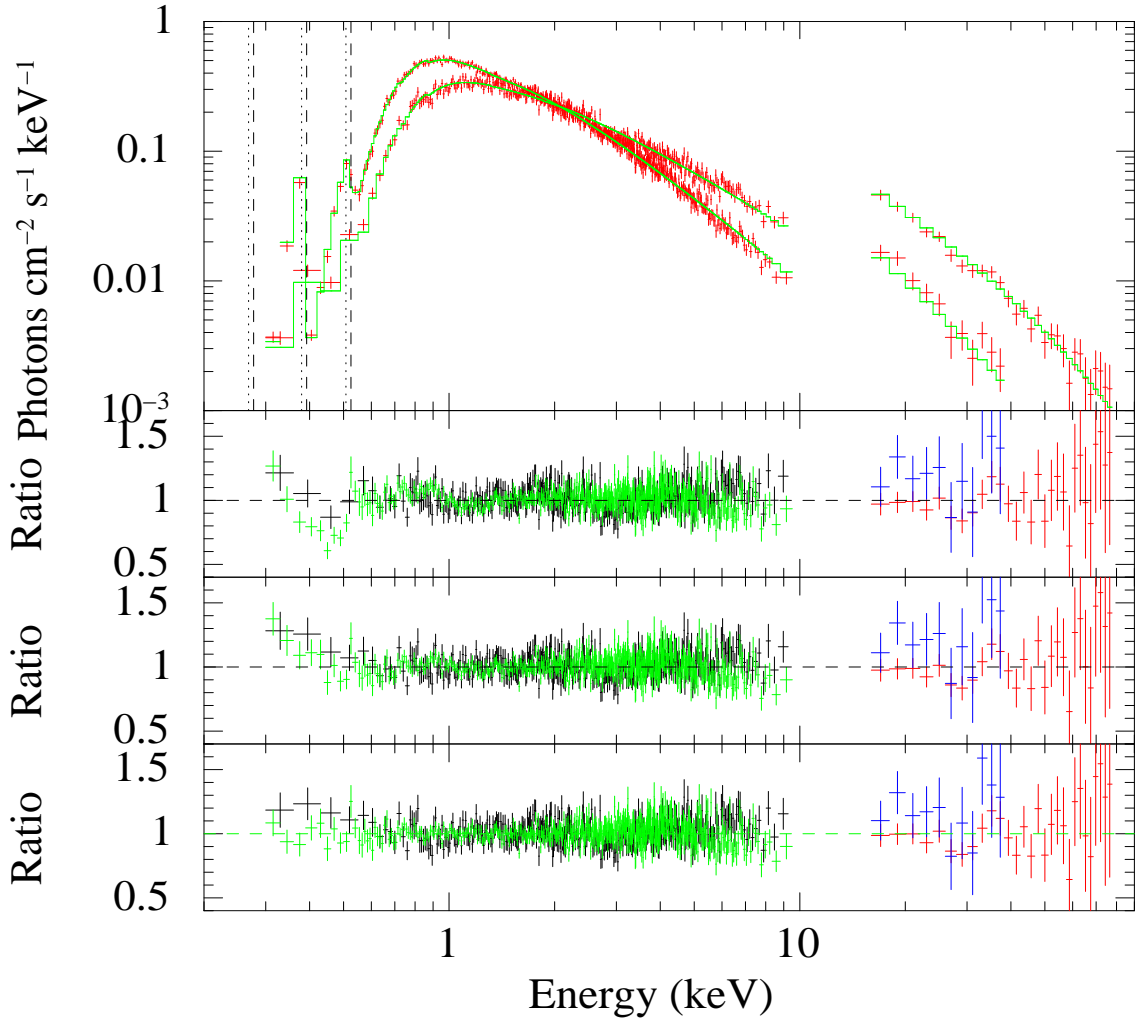}
\end{center}
\caption{The upper panel shows the XRT and BAT unfolded energy spectra for the
two time intervals discussed in the text. Dashed and dotted lines indicate the
rest-frame and redshifted edge energy of the CNO, respectively.
The three lowest panels show the spectral residuals as ratio between the data
and the following spectral model. For the second panel we considered a
solar metallicity absorber in the host galaxy. For the third panel we
let the metallicity of the host galaxy absorption free to vary
(keeping the standard solar abundance pattern fixed). For the lowest panel we  
fixed the the metallicity to the one derived from optical studies at
$0.07\,Z_\odot$ of the entire host galaxy and leave free only the CNO abundances
(for both absorbing models we use the {\tt tbvarbas} model).
}
\label{lc}
\end{figure*}

\begin{figure*}[htbp]
\begin{center}
\psfig{figure=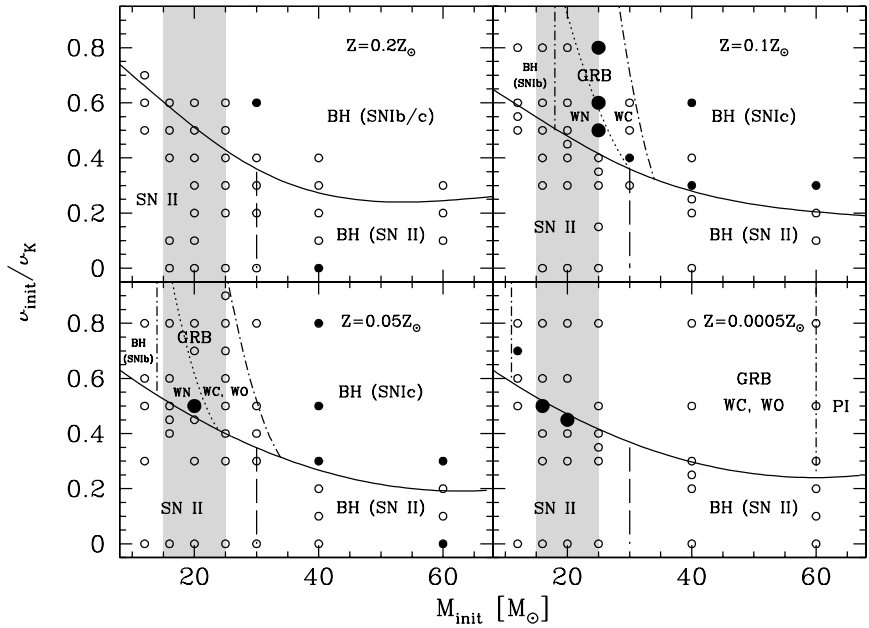,width=13truecm,angle=-90}
\end{center}
\caption{Constraints from the abundance ratio in the plane of initial mass and
initial stellar velocity (in units of equatorial Keplerian velocity) as
described in Yoon et al. (2006). The four panels refer to four different initial
metallicities ($0.2,\ 0.1,\ 0.05,\ 0.0005\,Z_\odot$ clockwise). We note that
the mean host galaxy metallicity has been estimated in $Z\sim 0.07\,Z_\odot$.
Different final end-products can be identified in this plane for
rotating single massive stars.
The solid line divides the plane into two parts,
where stars evolve quasi-chemically homogeneous above the line, while they
evolve into the classical core-envelope structure below the line. The
dotted-dashed lines bracket the region of quasi-homogeneous evolution where
the core mass, core spin and stellar radius are compatible with the collapsar
model for GRB production (absent at $Z=0.2\,Z_\odot$). The GRB production
region is divided into two parts, where GRB progenitors are WN or WC/WO types.
The dashed line in the region of non-homogeneous evolution separates Type II
supernovae (SN II; left) and black hole (BH; right) formation, where the
minimum mass for BH formation is simply assumed to be $30\msole$. We added to
these planes single star models allowed by the abundance constraints derived
from the X--ray spectrum of GRB060218. Filled circles represent single star
models from Yoon et al. (2006) which satisfy the abundance constraints derived 
from X--ray data; open circles represent instead models not satisfying these constraints.
Larger dots refer to models within the $15-25\msole$ (initial) progenitor mass range 
(i.e. within the vertical strip), based the mass estimate from Mazzali et al. (2006).} 
\label{lc}
\end{figure*}

\end{document}